\documentclass[%
    11pt,
    DIV=12,
    BCOR=0mm,
    twoside=semi,
    headings=small
]{scrartcl}

\usepackage{scrpage2}

\addtokomafont{title}{\normalfont \bfseries}
\addtokomafont{section}{\normalfont \bfseries}
\addtokomafont{subsection}{\normalfont \bfseries}
\addtokomafont{subsubsection}{\normalfont \bfseries}
\addtokomafont{paragraph}{\normalfont \bfseries}
\newcommand{\quoteparagraph}[1]{\noindent{\normalsize \bfseries #1}\hspace{1em}}

\usepackage[utf8]{inputenc}           % support for character encoding
\usepackage{amsmath, amsthm, amssymb} % pakages of American Mathematical Society (AMS)
\usepackage[english]{babel}           % support for English language
\usepackage{bibgerm}                  % support for german citation style
\usepackage{graphicx}                 % grafics (LaTeX: eps; pdfLaTeX: pdf, jpg, png)
\usepackage{hyperref}                 % automatic cross references
\usepackage{color}
\usepackage{subcaption}

\newtheorem{assumption}{Assumption}

\usepackage{bookmark}
\hypersetup{
  pdftitle    = {Backtest of Trading Systems on Candle Charts},
  pdfauthor   = {Stanislaus Maier-Paape and Andreas Platen},
  pdfkeywords = {backtest evaluation, historical simulation, trading system, candle chart, imperfect data},
  pdfborder={0 0 0.5}
}

\author{
  \normalsize \textsc{Stanislaus Maier-Paape}\\[-0.2em]
    \small \textit{Institut f\"ur Mathematik, RWTH Aachen,}\\[-0.5em]
    \small \textit{Templergraben 55, D-52052 Aachen, Germany}\\[-0.5em]
    \small \href{mailto:maier@instmath.rwth-aachen.de}{maier@instmath.rwth-aachen.de}\\
  \\[-0.75em]
  \normalsize \textsc{Andreas Platen}\\[-0.2em]
    \small \textit{Institut f\"ur Mathematik, RWTH Aachen,}\\[-0.5em]
    \small \textit{Templergraben 55, D-52052 Aachen, Germany}\\[-0.5em]
    \small \href{mailto:platen@instmath.rwth-aachen.de}{platen@instmath.rwth-aachen.de}
}
\date{
  \vspace{0.25em}
  \normalsize\today
  \vspace{-1cm}
}
\title{
  \vspace{-2cm}
  \Large Backtest of Trading Systems on Candle Charts
}

\clearscrheadfoot
\lehead{\pagemark}                   %% Top left on even pages
\rohead{\pagemark}                   %% Top right on odd pages
\begin{document}

\maketitle

\begin{quote}
  \small
  \quoteparagraph{Abstract}
  In this paper we try to design the necessary calculation needed for backtesting trading systems when only candle chart data are available. We lay particular emphasis on situations which are not or not uniquely decidable and give possible strategies to handle such situations.

  \quoteparagraph{Keywords} backtest evaluation, historical simulation, trading system, candle chart, imperfect data

  \quoteparagraph{JEL classification}
  % C: Mathematical and quantitative methods
  %% C1: Econometric and Statistical Methods: General
  C15, % Statistical Simulation Methods: General
  %% C6: Mathematical Methods; Programming Models; Mathematical and Simulation Modeling
  C63, % Computational techniques; Simulation modeling
  %% C9: Design of Experiments
  C99  % Other

\end{quote}

%%%%%%%%%%%%%%%%%%%%

\section{Introduction}

Since at least a decade more and more software solutions for self-designable trading systems have emerged (e.g. Ninjatrader, Tradestation, Tradesignal online, Nanotrader, Investox, to name just a few). All of the listed also incorporate a backtesting (also called historical simulation) tool including helpful statistical data on the trading success, i.e. it is possible to run a trading system on historical data to simulate the trades. The idea is that trading systems which were successful in the past should be successful in the future as well. Analogously a trading system which performs bad on historical data cannot be trusted and is supposed to be unsuccessful in the future. This makes backtesting an important tool for designing trading systems.

Although already several years on the market we found that many of the software solutions perform calculations sometimes incorrectly. This concerns even situations which are uniquely decidable. When backtests are evaluated just on the knowledge of candle data, however, there are always situations, which can not or not uniquely (SNU: situation which is not unique) be determined, see e.g. the book of Pardo~\cite[Chapter 6, Section ``Software Limitations'']{Pardo2008} or Harris~\cite[Chapter 6]{Harris2008}. Pardo~\cite{Pardo2008} and Harris~\cite{Harris2008} describe this problem but do not discuss the backtest algorithm itself and how to deal with such problems. The least what a backtest engine should do in these situations is to warn the user about these problems. Also the user should be informed, how such situations are handled. We suggest that there should be four different strategies to choose from:
\vspace*{0.2cm}

\begin{tabular}{r@{\hspace{0.15cm}}r@{\hspace{0.2cm}}l}
  {\bfseries I.}   & worst case (wc): & the SNU is evaluated as the worst possible case for the user.     \\ [0.25cm]
  {\bfseries II.}  &  best case (bc): & the SNU is evaluated as the best possible case for the user.      \\ [0.25cm]
  {\bfseries III.} &     ignore (ig): & the entry signal or the whole trade is ignored.                   \\ [0.25cm]
  {\bfseries IV.}  &      exact (ex): & to resolve the problem, more data (sometimes even tick data)      \\
             &                  &   have to be loaded.
\end{tabular}
\vspace{2mm}

To the best of the author's knowledge there is no publication about backtest algorithms itself but only for the statical evaluation of backtests. Typical statistical measures like Sharpe ratio, average trade, profit factor and many more, see e.g. \cite[Chapter 22]{KD2011}, give hints on how the trading system performs.

Therefore we discuss the procedure of backtest evaluation based on candle/bar chart data in detail. Further information about backtesting and some limitations can be found e.g. in the books of Chan~\cite[Chapter 3]{Chan2009}, Pardo~\cite[Chapter 6]{Pardo2008} and Harris~\cite[Chapter 6]{Harris2008} and for trading options in the book of Izraylevich and Tsudikman~\cite[Chapter 5]{IT2012}.

It is well known that a backtest is just a simulation over the past and does not predict future behavior of a trading system. The ability to accurately simulate a parameter dependent trading system on some chart data can rapidly lead to an overestimation of the parameters by optimizing these parameters to reach the best performance on the historical data. Ni and Zhang~\cite{NZ2005} present a method to improve the efficiency of backtesting a trading strategy for different parameter choices but they do not explain the backtest evaluation itself. The result could be an optimal trading system but only well adjusted to the past. In general this does not mean that this parameter setting is also appropriate in the future and gives a stable strategy. In contrast this can lead to tremendous losses. This phenomena is called backtest overfitting, see \cite{BBP+2014,BBP+2014a,CP2014} and also \cite[Chapter 6]{Pardo1992} for a detailed discussion. Therefore backtesting needs to be used carefully but, nevertheless, gives important information about a trading strategy.

Clearly the above remarks and references show that a correct interpretation of backtest results is a difficult and more or less up to now unsolved problem. However, this is not the subject of our considerations in this paper. Here we want to focus the attention on how the backtest evaluation itself has to be calculated correctly.

Due to symmetry, it suffices to consider entry orders for long positions only. Therefore we discuss only long positions unless we explicitly refer to short orders. Since market orders are to be executed at the open of the next candle, problems of backtest evaluation for the position entry only occur for ``limit buy'' (with limit level $l^*$), ``stop buy'' (with stop level $b^*$) and ``stop limit buy'' (with stop level $b^*$ and limit level $l^*$) long orders, see e.g. \cite[Chapter 4]{Pardo1992} for definition of some order types.

We discuss the principal part of this paper, i.e. the decisions for backtest evaluation, in Section~\ref{sec:backtest_alg}, while in Subsection~\ref{sec:assumptions} we need to make some assumptions and discuss some limitations of a backtest. In Subsections~\ref{sec:limit} to \ref{sec:stop_limit} we discuss the question when and how a position has to be opened with the classical ``EnterLongLimit()'', ``EnterLongStop()'' and ``EnterLongStop\-Limit()'' orders, respectively. In all three cases the decision tree is only given for the first bar of the trade. Since typically a trading setup includes immediate stop losses (at $s^*$) or target levels (at $t^*$), even in the first bar besides the pure position entry, there are numerous other things to check. Once the first bar of the trade has finished or in case we enter the position immediately at the beginning of the period the decisions for such an active position in succeeding bars is simpler. The decision tree for the latter is given in Subsection~\ref{sec:exit}.
We close the discussion with the conclusion in Section~\ref{sec:conclusions}.

%%%%%%%%%%%%%%%%%%%%

\section{Backtest evaluation algorithm}\label{sec:backtest_alg}

We look at situations for different entry and exit setups. All orders are generated at the end of a candle so that these orders can be filled in the next candle. Therefore we take a look at this next candle for different orders.
The examined candle has the four values $H=$ High, $L=$ Low, $O=$ Open and $C=$ Close.

%%%%%%

\subsection{Assumptions and limitations}\label{sec:assumptions}

In order to be able to perform exact calculations we first need to make a continuity assumption on the price evolution within a candle.

\begin{assumption}\label{ass:nipg}
  (No intra-period gaps)\\
  We assume that the price evolution inside the period skips no nearby tick-values, i.e. starting at the open until the end of the period at the close all price moves during that period (up or down) come only as  $\pm$\,1 tick. Intra-period gaps, i.e. moves by more than one tick, thus are not allowed.
\end{assumption}

This assumptions is essential for determining intra-period entry or exit prices, e.g. at limit or stop levels. In live trading, however, this assumption is not realistic. To overcome this problem usually slippage is introduced for each backtest trade, see e.g. the book of Pardo~\cite[Chapter 6, Section ``Realistic Assumptions'']{Pardo2008} for a detailed discussion.

Additionally we need to assume that all orders are filled at the requested price.

\begin{assumption}
  (Market liquidity)\\
  We assume that we trade on a perfectly liquid market. I.e. our orders do not affect the price changes and are fully filled at the corresponding entry or stop level.
\end{assumption}

Of course this assumption is also not realistic. Similar to Assumption~\ref{ass:nipg} slippage can help to get more reasonable results.

Since all prices (measured as tick-values) are integers we need to make sure that all values given by the user (like limit level $l^*$, etc.) are of the same type to avoid rounding errors, see also \cite[Chapter 6, Section ``Software Limitations'']{Pardo2008}.

\begin{assumption}
  (Rounded values)\\
  All values like limit price, target and stop loss level are given as numbers which are rounded to the corresponding next possible price value which depends on the tick size.

  For long positions the stop level $b^*$ for stop buy order and the target level $t^*$ are rounded up while the stop loss level $s^*$ and limit price $l^*$ are rounded down.

  For short positions the stop level $b^*$ for stop buy order and the target level $t^*$ are then rounded down while the stop loss level $s^*$ and limit price $l^*$ needs to be rounded up.
\end{assumption}

This way to round the values does not change any decision during the backtest evaluation but it corrects the price values used for the computation of the outcomes of each trade.

In case the long position is coupled with a stop loss order (stop level $s^*$) or a target order (target level $t^*$) we always assume $s^*< l^*< t^*$ and $s^*< b^* < t^*$.

Next we need to make some simplification for the best and worst case for SNUs. Suppose we are invested in a stock and there is a SNU with the two options of exit the position at target level $t^*$ or to stay invested. Of course exit at $t^*$ should immediately lead to a win trade. However, if we change the target in the next period it is possible to earn even more money if we do not exit the position at this moment but later in one of the subsequent candles. This can also affect upcoming trades which in general depend on the current status (invested or not invested) and thus can increase the complexity of the decisions needed to be done for the real (globally) best case. Therefore we always choose the simplest setting which is best for the user at the current period. In this easy example this would be to immediately exit the position at target level $t^*$.

\begin{assumption}\label{ass:wc_bc}
  (Worst and best case)\\
  Best and worst case decisions in situations which cannot be uniquely determined (SNU) should be made on premise that it is best/worst for the current period only.
\end{assumption}

%%%%%%

\subsection{Entry of a long position with ``limit buy'' order}\label{sec:limit}
\begin{figure}[bht]
  \centering
  \begin{subfigure}[b]{54.57mm}
    \centering
    \includegraphics{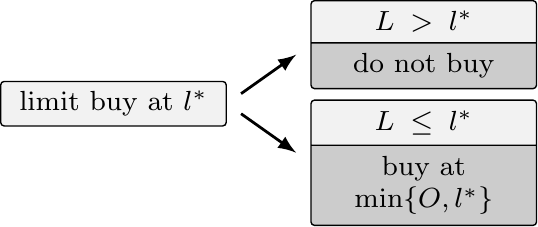}
    \caption{Only limit buy long order.}
    \label{fig:limitbuy}
  \end{subfigure}
  \hfill
  \begin{subfigure}[b]{86.1mm}
    \centering
    \includegraphics{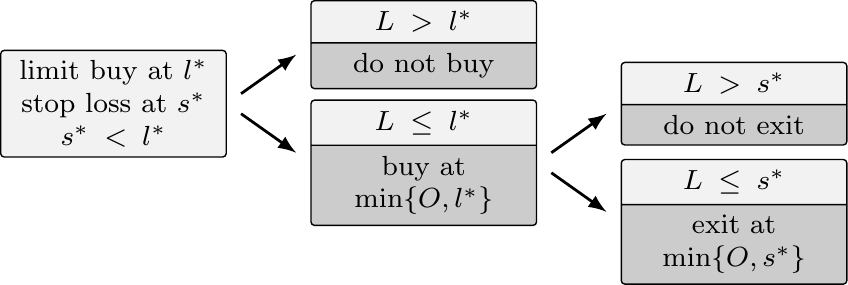}
    \caption{Limit buy long order supplemented with stop loss.}
    \label{fig:limitbuy_stoploss}
  \end{subfigure}
  \caption{Entry setups with limit buy long order.}
  \label{fig:limitbuy_split}
\end{figure}

\begin{figure}[bht]
  \centering
  \includegraphics[scale=0.99]{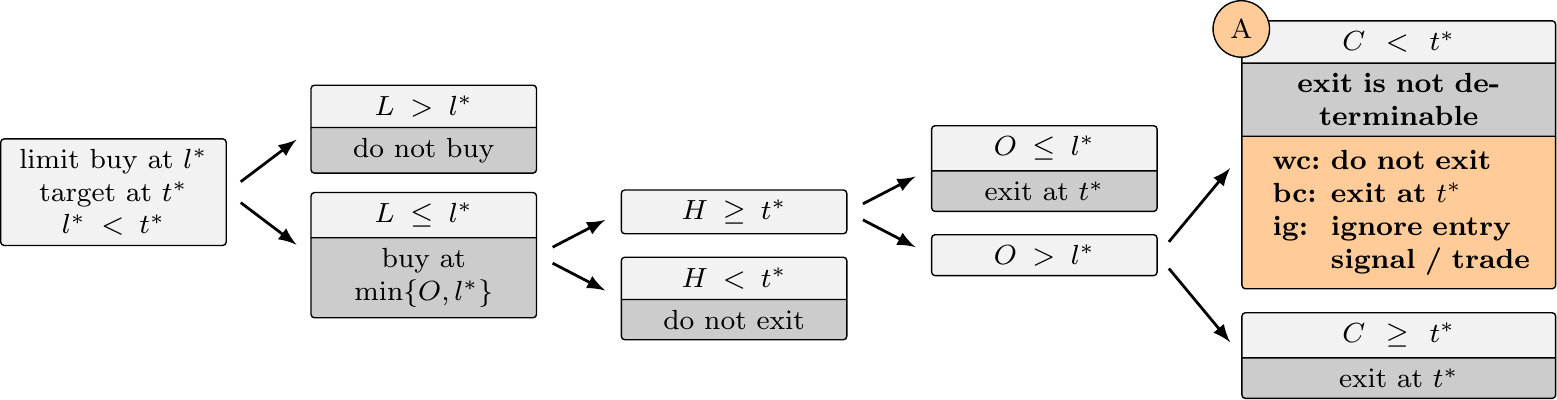}
  \caption{Entry setup with limit buy long order supplemented with target.}
  \label{fig:limitbuy_target}
\end{figure}

\begin{figure}
  \centering
  \includegraphics{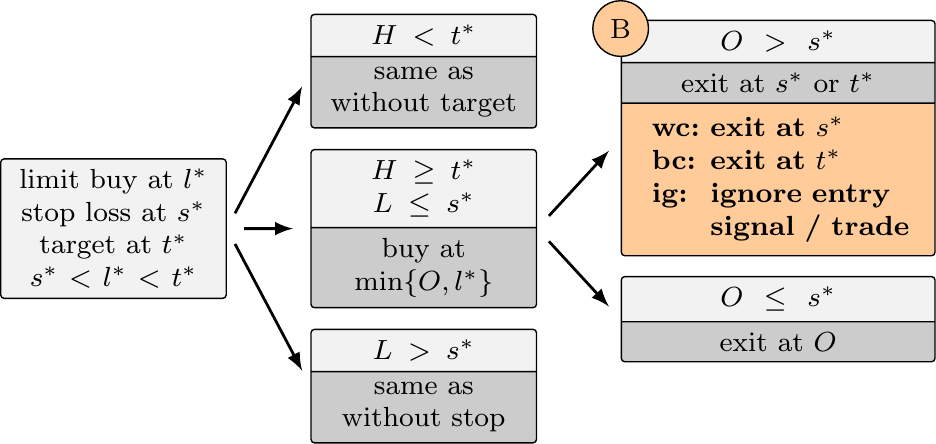}
  \caption{Entry setup with limit buy long order supplemented with stop loss and target.}
  \label{fig:limitbuy_stoploss_target}
\end{figure}

Here the long position is only opened, once the price reaches the limit level $l^*$ (or below). This is the situation of an entry with the classical ``EnterLongLimit()'' order optionally supplemented by stop loss $s^*$ and target levels $t^*$. We assume $s^*<l^*<t^*$. The decision trees are shown in Figures~\ref{fig:limitbuy_split} to \ref{fig:limitbuy_stoploss_target}.

\pagebreak

\begin{figure}
  \centering
  \begin{subfigure}[b]{0.49\linewidth}
    \centering
    \includegraphics{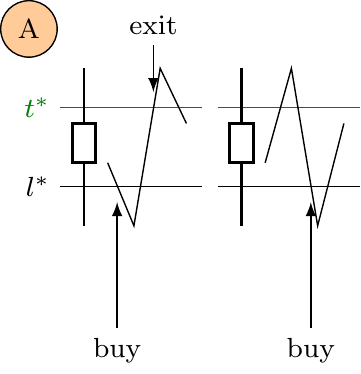}
    \caption{Cases for Figure~\ref{fig:limitbuy_target}.}
  \end{subfigure}
  \hfill
  \begin{subfigure}[b]{0.49\linewidth}
    \centering
    \includegraphics{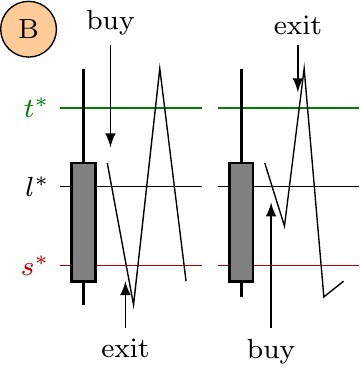}
    \caption{Cases for Figure~\ref{fig:limitbuy_stoploss_target}.}
  \end{subfigure}
  \caption{Variations of possible price development within a candle for SNU with limit buy long order.}
  \label{fig:limitbuy_cases}
\end{figure}

In Figures~\ref{fig:limitbuy_target} and \ref{fig:limitbuy_stoploss_target} the cases marked with A and B, respectively, are two SNUs, i.e. if we cannot load extra data like tick data to make these situations unique, there are multiple possibilities for the correct position entry and/or exit. Figure~\ref{fig:limitbuy_cases} shows one example for each possibility for both SNUs A and B.

We always assume $s^*< l^*$ because of the following reason: In case $s^*\geq l^*$ the position would be closed right after it is opened, which makes no sense and should therefore be forbidden by the software, i.e. these order should be canceled/ignored.

If $t^* \leq l^*$ the same would happen if $O\geq t^*$ and of course $L\leq l^*$. However, if $O<t^*$ the position would be opened at the beginning of the period which is equivalent to a market order executed at the open of the subsequent period. In this case the trade would not immediately be stopped and thus can be handled as in Subsection~\ref{sec:exit} if it is not ignored in advance.

From the decision trees in Figures~\ref{fig:limitbuy_split} to \ref{fig:limitbuy_stoploss_target} we see that for limit orders only a combination involving a target where the target is reached in the entry period leads to SNUs. If there is no target or if the target is far away all situations are uniquely decidable.

%%%%%%

\subsection{Entry of a long position with ``stop buy'' order}\label{sec:stop}
\begin{figure}
  \centering
  \begin{subfigure}[b]{54.57mm}
    \centering
    \includegraphics{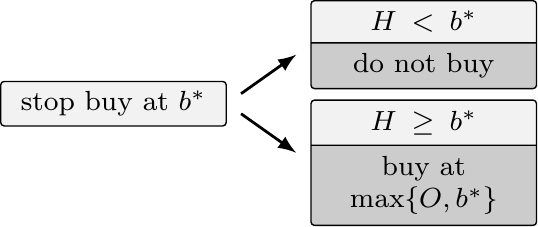}
    \caption{Only stop buy long order.}
    \label{fig:stopbuy}
  \end{subfigure}
  \hfill
  \begin{subfigure}[b]{86.1mm}
    \centering
    \includegraphics{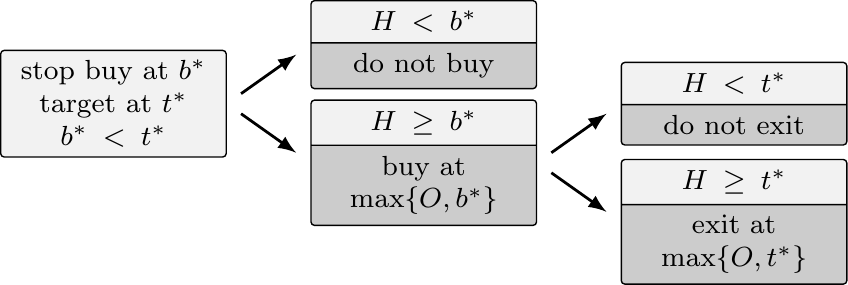}
    \caption{Stop buy long order supplemented with target.}
    \label{fig:stopbuy_target}
  \end{subfigure}
  \caption{Entry setups with stop buy long order.}
  \label{fig:stopbuy_split}
\end{figure}

\begin{figure}
  \centering
  \includegraphics[scale=0.99]{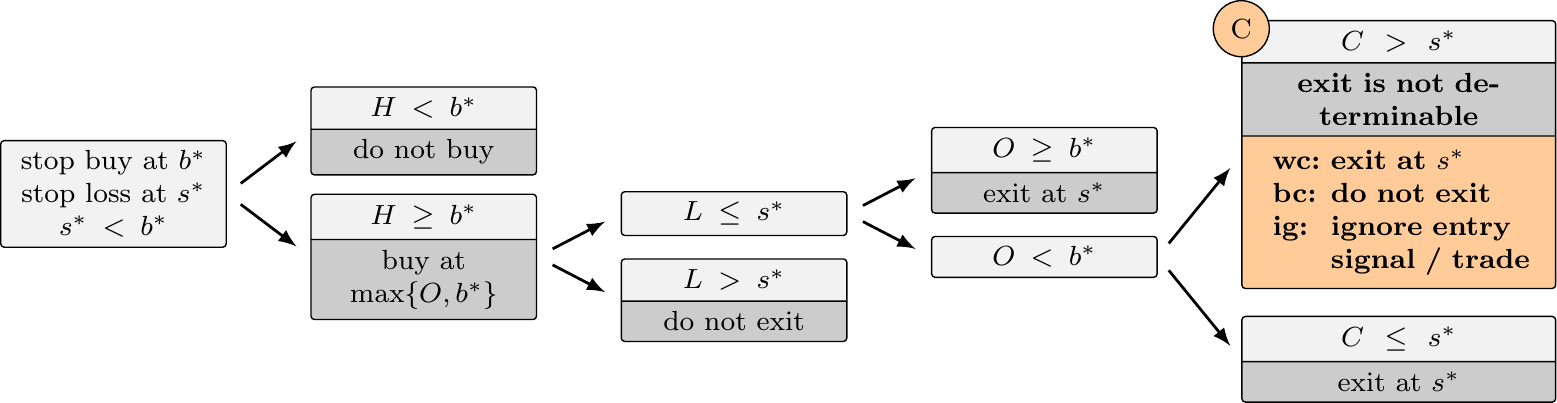}
  \caption{Entry setup with stop buy long order supplemented with stop loss.}
  \label{fig:stopbuy_stop}
\end{figure}

\begin{figure}
  \centering
  \includegraphics{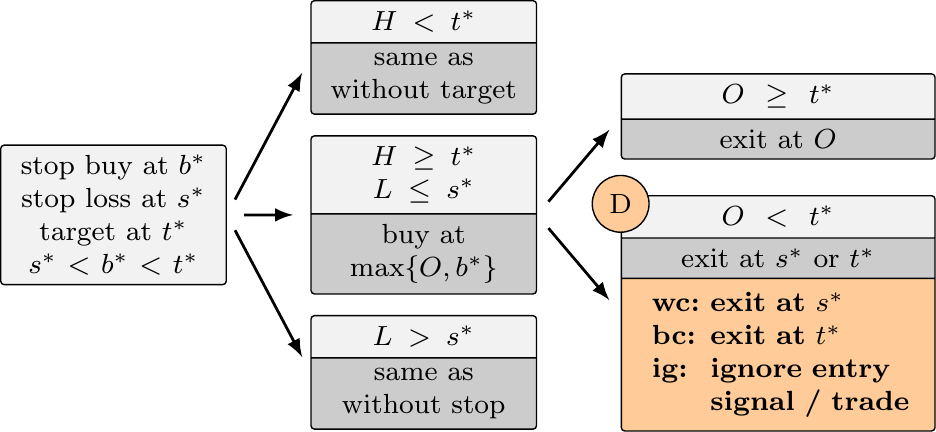}
  \caption{Entry setup with stop buy long order supplemented with stop loss and target.}
  \label{fig:stopbuy_stop_target}
\end{figure}

\begin{figure}
  \centering
  \begin{subfigure}[b]{0.49\linewidth}
    \centering
    \includegraphics{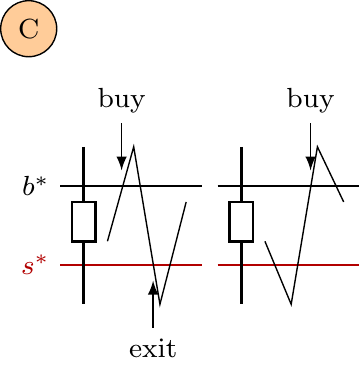}
    \caption{Cases for Figure~\ref{fig:stopbuy_stop}.}
  \end{subfigure}
  \hfill
  \begin{subfigure}[b]{0.49\linewidth}
    \centering
    \includegraphics{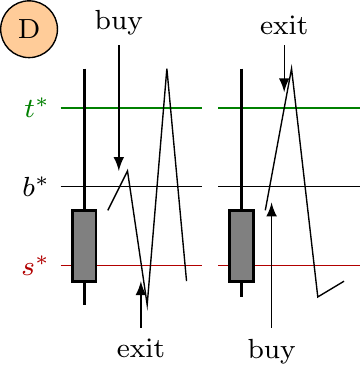}
    \caption{Cases for Figure~\ref{fig:stopbuy_stop_target}.}
  \end{subfigure}
  \caption{Variations of possible price development within a candle for SNU with stop buy long order.}
  \label{fig:stopbuy_cases}
\end{figure}

Here the long position is only opened, once the price reaches the stop level $b^*$ (or above), created by the classical ``EnterLongStop()'' order. Again the order can optionally be supplemented by stop loss ($s^*$) and target levels ($t^*$) with $s^*<b^*<t^*$. The decision trees are shown in Figures~\ref{fig:stopbuy_split} to \ref{fig:stopbuy_stop_target} and the examples for the SNUs in Figure~\ref{fig:stopbuy_cases}.

Since an entry stop order is some kind of mirrored version of the entry limit order, we now have SNUs for the entry stop order supplemented with an initial stop loss level.

Again, the case $t^*\leq b^*$ makes no sense, because the position would be closed immediately after the opening, compare the case $s^*\geq l^*$ for long limit order.

In the case $b^*\leq s^*$ the position is either to be closed after opening if $O\leq b^*$ or, if $O> b^*$, we have an equivalent case to a market order.

%%%%%%

\subsection{Entry of a long position with ``stop limit buy'' order}\label{sec:stop_limit}
\begin{figure}[bht]
  \centering
  \includegraphics{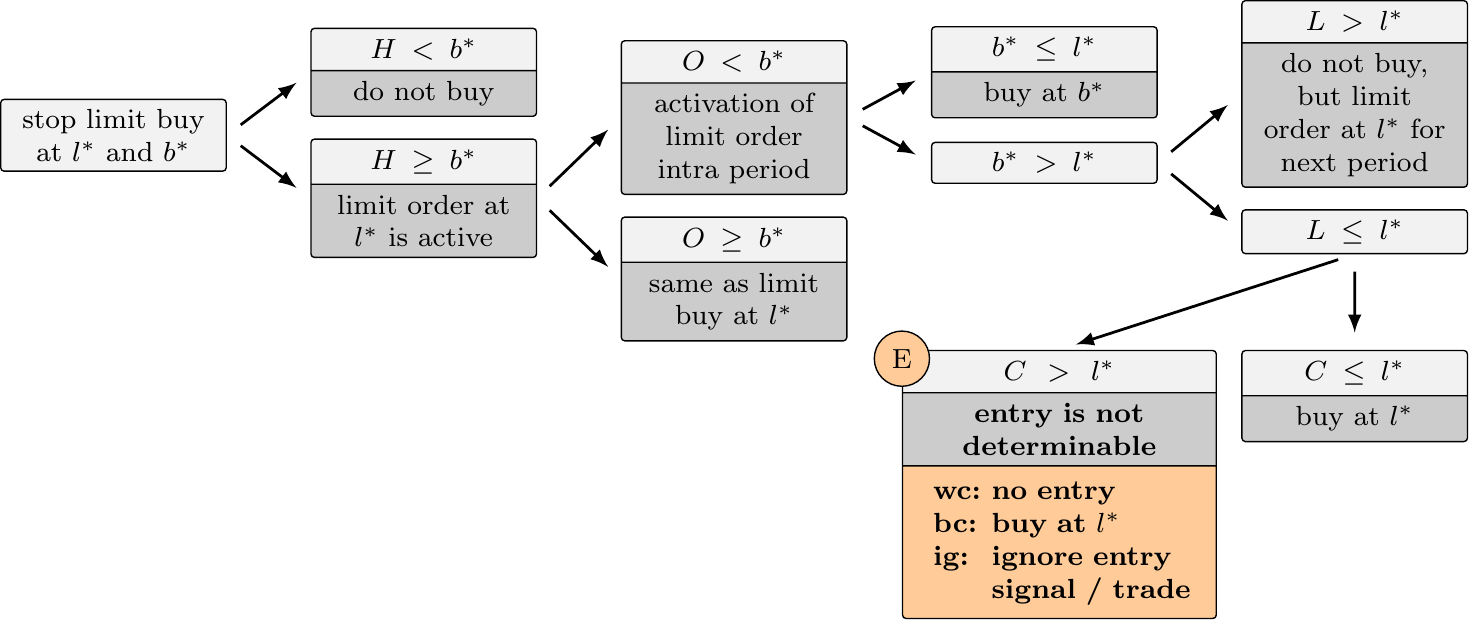}
  \caption{Entry setup with stop limit buy long order.}
  \label{fig:stoplimitbuy}
\end{figure}

\begin{figure}
  \centering
  \includegraphics[scale=0.99]{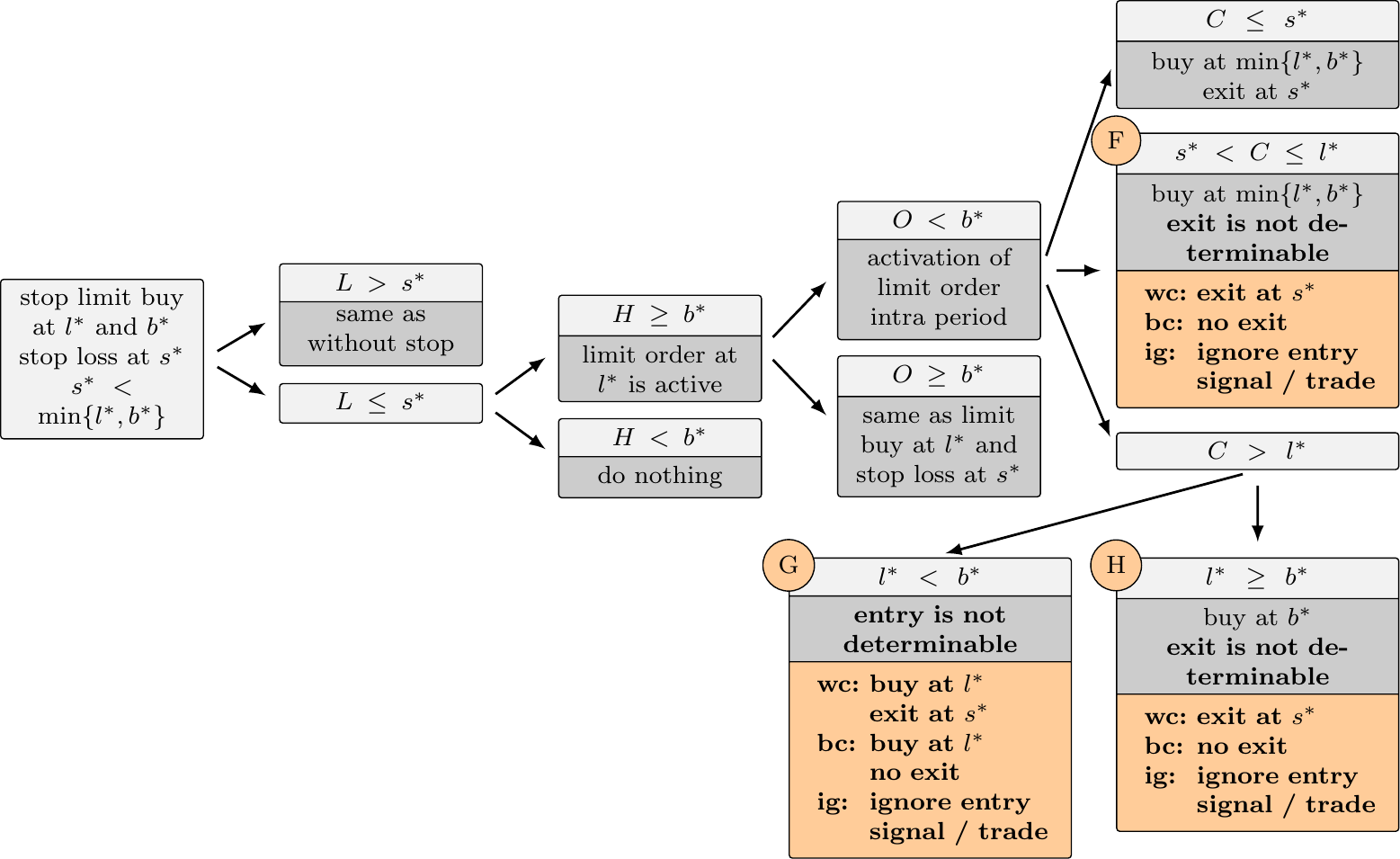}
  \caption{Entry setup with stop limit buy long order supplemented with stop loss.}
  \label{fig:stoplimitbuy_stop}
\end{figure}

\begin{figure}
  \centering
  \includegraphics{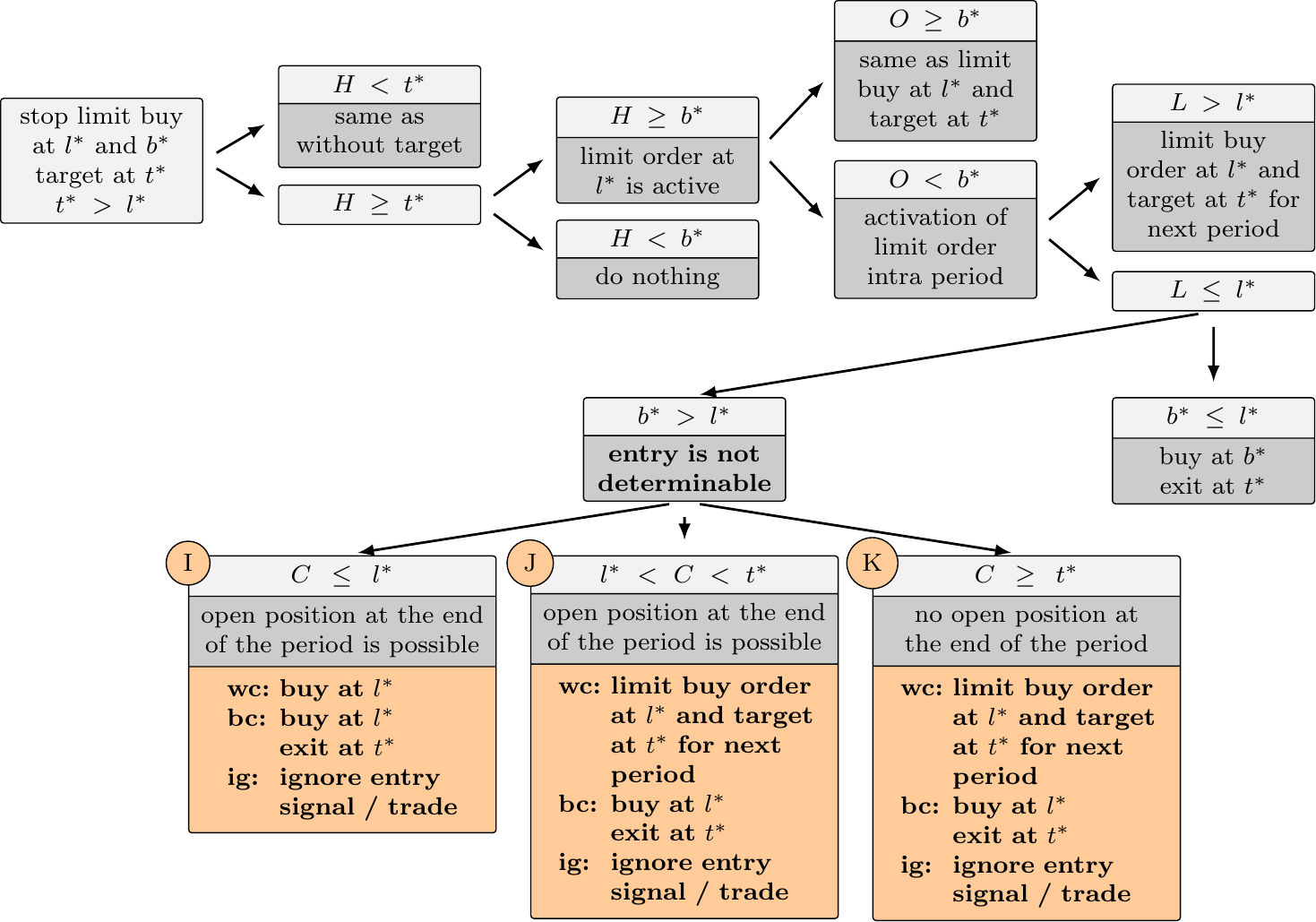}
  \caption{Entry setup with stop limit buy long order supplemented with target.}
  \label{fig:stoplimitbuy_target}
\end{figure}

\begin{figure}
  \centering
  \includegraphics[scale=0.99]{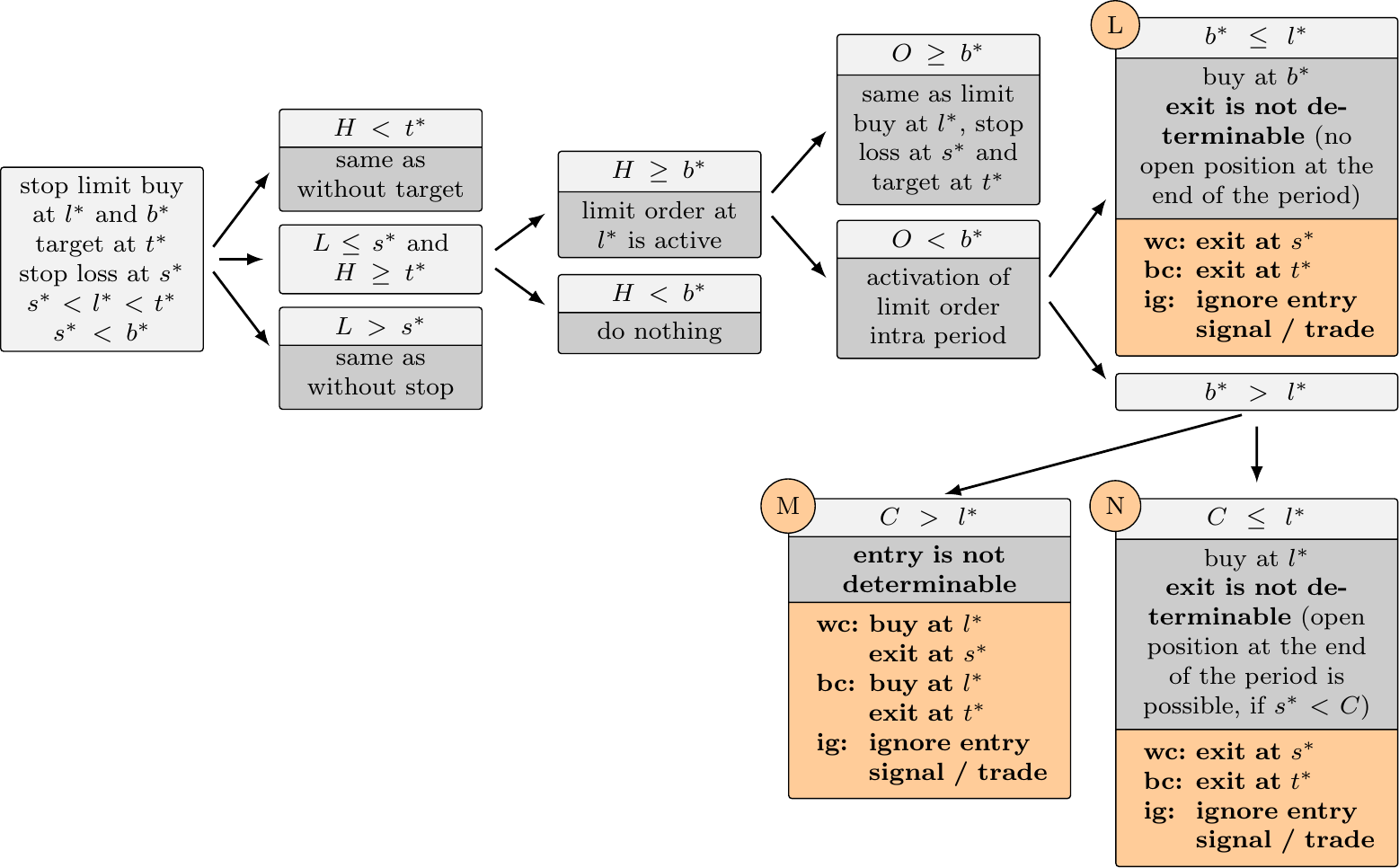}
  \caption{Entry setup with stop limit buy long order supplemented with stop loss and target.}
  \label{fig:stoplimitbuy_stop_target}
\end{figure}

\begin{figure}
  \centering
  \includegraphics{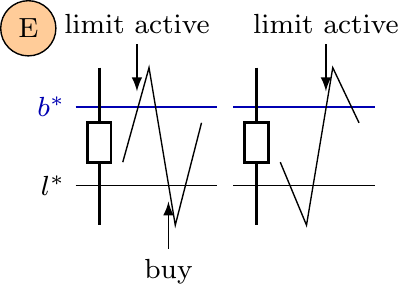}
  \caption{Variations of possible price development within a candle for SNU with stop limit buy long order for Figure~\ref{fig:stoplimitbuy}.}
  \label{fig:stoplimitbuy_cases}
\end{figure}

\begin{figure}
  \centering
  \includegraphics{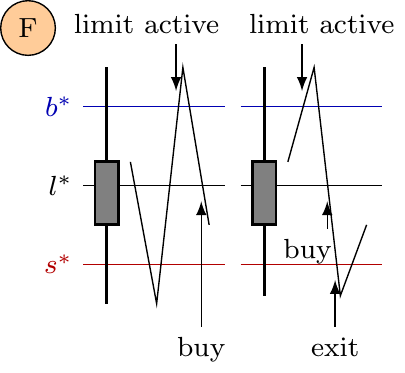}
  \hfill
  \includegraphics{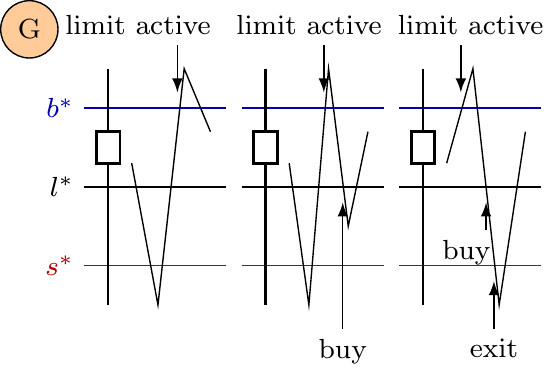}
  \hfill
  \includegraphics{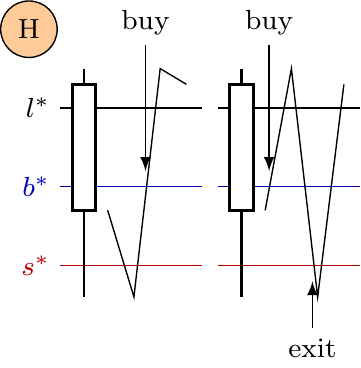}
  \caption{Variations of possible price development within a candle for SNU with stop limit buy long order for Figure~\ref{fig:stoplimitbuy_stop}.}
  \label{fig:stoplimitbuy_cases_stop}
\end{figure}

\begin{figure}
  \centering
  \includegraphics{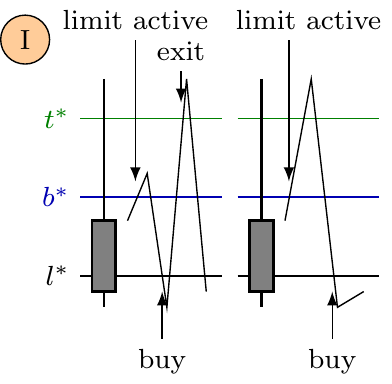}
  \hfill
  \includegraphics{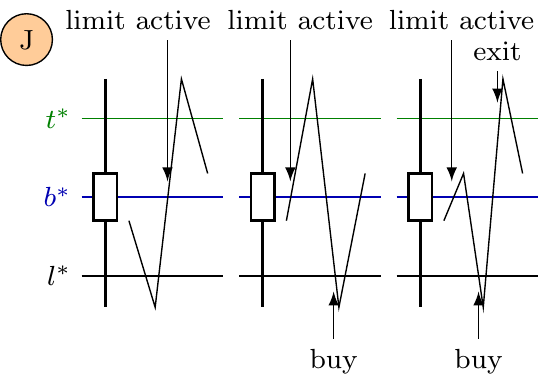}
  \hfill
  \includegraphics{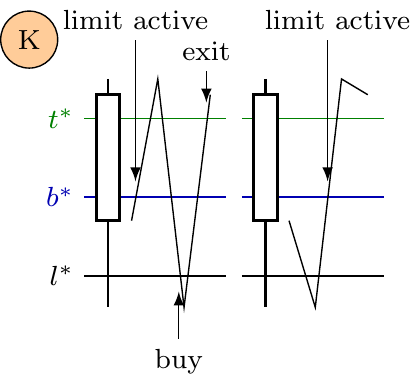}
  \caption{Variations of possible price development within a candle for SNU with stop limit buy long order for Figure~\ref{fig:stoplimitbuy_target}.}
  \label{fig:stoplimitbuy_cases_target}
\end{figure}

\begin{figure}
  \centering
  \includegraphics[scale=0.9]{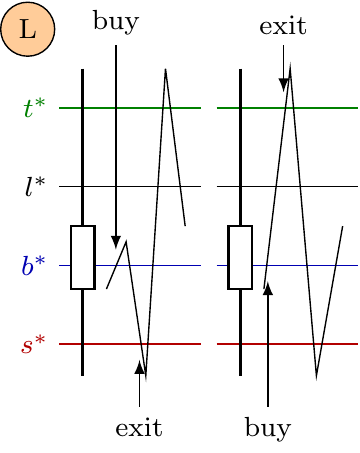}
  \hfill
  \includegraphics[scale=0.9]{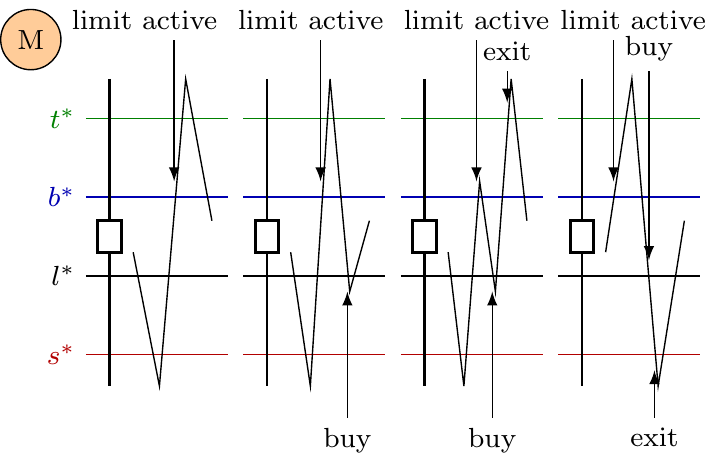}
  \hfill
  \includegraphics[scale=0.9]{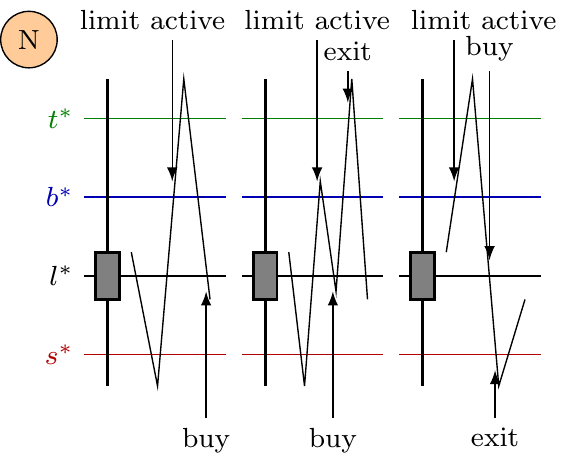}
  \caption{Variations of possible price development within a candle for SNU with stop limit buy long order for Figure~\ref{fig:stoplimitbuy_stop_target}.}
  \label{fig:stoplimitbuy_cases_stop_target}
\end{figure}

Here a limit buy order at level $l^*$ is only generated, once the price reaches the stop level $b^*$ (or above), as is generated by the classical ``EnterLongStopLimit()'' order. I.e. the trader in principle wishes to have an ``EnterLongLimit()'' order at level $l^*$, but to activate that order he firstly wants that the prices reach the stop level $b^*$ (or higher). Again this order may optionally be supplemented by stop loss ($s^*$) or target levels ($t^*$). We assume $s^*<\min\{l^*,b^*\}\leq l^* < t^*$. The decision trees are shown in Figures~\ref{fig:stoplimitbuy} to \ref{fig:stoplimitbuy_stop_target}, and examples for the SNUs in Figures~\ref{fig:stoplimitbuy_cases} to \ref{fig:stoplimitbuy_cases_stop_target}, respectively.

This entry order type is much more complex and therefore leads to larger decision trees and much more SNUs. Even the worst cases and/or best cases for some SNUs are not uniquely determinable because there are situations where a position can be opened or there is just an active limit order at the end of the candle, see e.g. SNU E. In general it is not clear whether it is worst or best to have an active limit order or an open position at the end of the candle in such cases. Because of Assumption~\ref{ass:wc_bc} we decide to measure the quality of an open trade by the current value of the trade which in this case is the difference between the close of the candle and the entry price of the position. If the close is larger than the entry price we currently are in a positive trade (and thus the best case) which is better than having just an active limit order (worst case), and vice versa if the close is below the entry price.

%%%%%%
\pagebreak

\subsection{Exit from an active long position}\label{sec:exit}
\begin{figure}
  \centering
  \begin{subfigure}[b]{0.49\linewidth}
    \centering
    \includegraphics{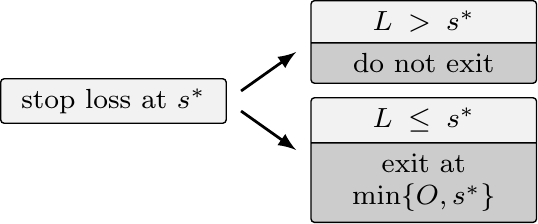}
    \caption{Only stop loss.}
  \end{subfigure}
  \begin{subfigure}[b]{0.49\linewidth}
    \centering
    \includegraphics{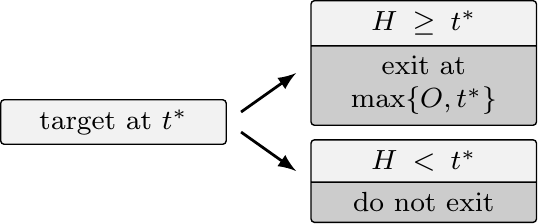}
    \caption{Only target.}
  \end{subfigure}
  \caption{Exit setups during an active long position.}
  \label{fig:exit}
\end{figure}

\begin{figure}
  \centering
  \includegraphics{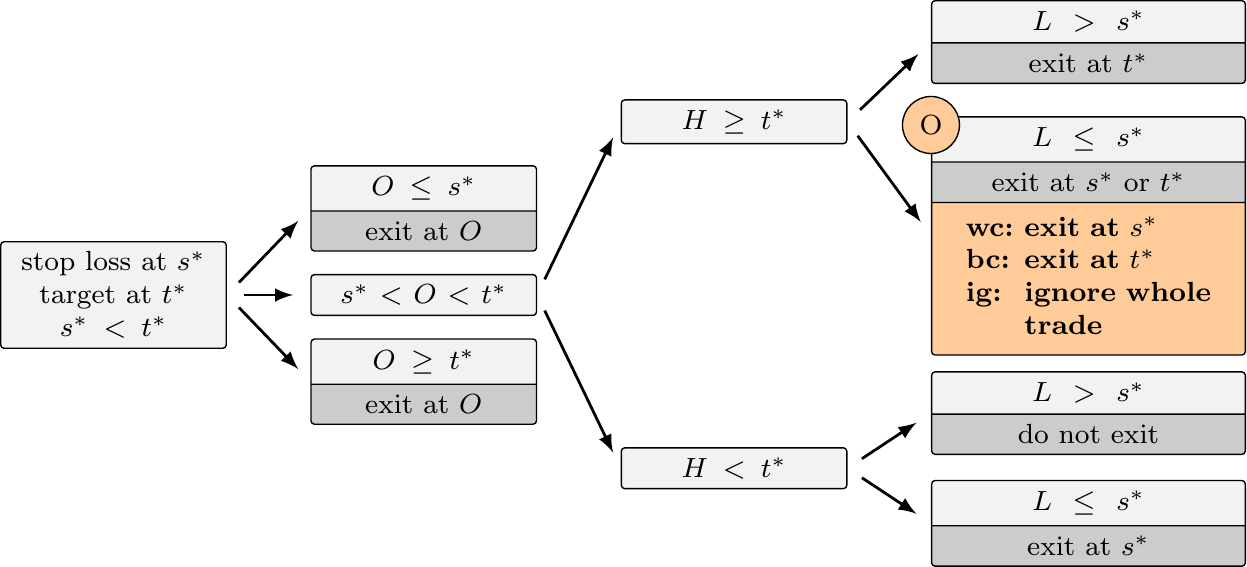}
  \caption{Exit setup during an active long position with both, stop loss and target.}
  \label{fig:exit_stop_target}
\end{figure}

\begin{figure}
  \centering
  \includegraphics{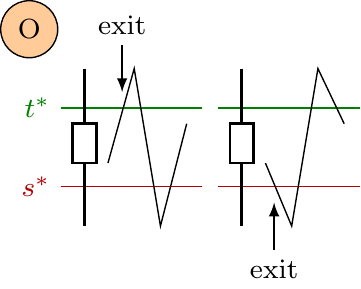}
  \caption{Variations of possible price development within a candle for SNU during an active long position for Figure~\ref{fig:exit_stop_target}.}
  \label{fig:exit_cases}
\end{figure}

The final discussion deals with the case of an active long position, i.e. at the end of the prior period a long position remained open. This also includes situations where a market order was generated in the prior candle such that a long position is opened right at the open of the current period. The decision trees for the current period are shown in Figures~\ref{fig:exit} and \ref{fig:exit_stop_target} and the examples for the SNUs in Figure~\ref{fig:exit_cases}.

%%%%%%%%%%%%%%%%%%%%
\pagebreak

\section{Conclusions}\label{sec:conclusions}

The precise listing of the backtest evaluation algorithm in the preceding section shows very clearly that not uniquely decidable situations (SNUs) are omnipresent when only candle data are available. This is not consistent with the fact that wide spread software solutions ignore that problem completely. An honest evaluation should give users the choice of worst/best case calculations. Future software solutions should be able to reload finer candle or tick data for the bars in question in order to evaluate backtests exactly.

%%%%%%%%%%%%%%%%%%%%

\end{document}